\documentstyle[aps,prl,epsfig,twocolumn]{revtex}

\begin{document}
\twocolumn[\hsize\textwidth\columnwidth\hsize\csname
 @twocolumnfalse\endcsname
\title{Hartree-Fock Theory of Hole Stripe States}
\draft
\author{Tae-Suk Kim$^{a}$ , S.R. Eric Yang$^{b}$, and A.H. MacDonald$^{c}$}
\address{$^a$ APCTP, 207-43 Cheongryangri-dong, Dongdaemun-gu, Seoul 130-012, Korea\\
 $^b$ Department of Physics, Korea University, Seoul, Korea\\
 $^c$ Department of Physics, Indiana University, Bloomington IN 47405}
\date{\today}
\maketitle
\begin{abstract}
We report on Hartree-Fock theory results for stripe states of two-dimensional
hole systems in quantum wells grown on GaAs (311)A substrates. We find that the
stripe orientation energy has a rich dependence on hole density, and on
in-plane field magnitude and orientation. Unlike the electron case, the
orientation energy is non-zero for zero in-plane field, and the ground state
orientation can be either parallel or perpendicular to a finite in-plane field.
We predict an orientation reversal transition in in-plane fields applied along
the $\lbrack\bar{2}33\rbrack$ direction.
\end{abstract}
\vskip1pc]

Because of the macroscopic dengeneracy of Landau levels, the physics of
two-dimensional (2D) electron systems in strong external fields has been a
fertile area for many-particle physics. Recently\cite{els1,du,els2,lilly2} the
emergence of strongly anisotropic transport properties at low temperatures has
been interpreted as evidence for the occurrence of the unidirectional
charge-density-wave {\em stripe} states predicted by Hartree-Fock
theory\cite{cdw}. For conduction band Landau levels with orbital kinetic energy
index $n>2$ (filling factor $\nu > 4$), the putative stripe states occur
instead of the strongly correlated fluid states responsible for the quantum
Hall effect\cite{leshouches}. Although Hartree-Fock theory provides a clear
motivation for stripe states\cite{cdw}, it cannot reliably predict the nature
of the ground state because the energetic competition with fluid states is
delicate\cite{rezayi}. Moreover, the transport properties of stripe states
cannot be explained by Hartree-Fock theory, although they are
consistent\cite{jpe00} with theories\cite{fertig,kivelson,ahm} of
quantum-fluctuating stripes.  For these reasons, the ability of Hartree-Fock
theory to predict\cite{jungwirth,phillips} the low resistance (parallel to
stripe) direction in an in-plane magnetic field has played an essential role in
establishing the stripe-state explanation of $N > 2$ anisotropic transport. A
recent study\cite{pan} in which a reorientation transition in a wide quantum
well sample is explained by Hartree-Fock theory is especially convincing in
this respect.

The present work is motivated by the discovery\cite{hole} of anisotropic
transport in 2D hole systems grown on GaAs (311)A substrates.  In this case,
anisotropic transport already occurs for $\nu \sim 5/2 $, demonstrating that
there are important differences between the electron and hole cases.  The
change is not unexpected, given the anisotropy of 2D hole band structure. We
have generalized the Hartree-Fock theory of stripe states to the case of
valence bands described by a the single-particle Luttinger
Hamiltonian\cite{lutham}. We find that strong orbital-quantum-number mixing
leads to anisotropic effective electron-electron interactions and to a
dependence of stripe energy on orientation even in the absence of an in-plane
magnetic field. This property favors the formation of a stripe state,
consistent with experiment. The ground state orientation is not in general
either parallel or perpendicular to the direction of a finite in-plane field
when one is present.

To describe 2D hole gases grown along the $[311]$ direction, it is convenient
to choose a Cartesian coordinate system with $[01\bar{1}]$, $[\bar{2}33]$ and
$[311]$ direction axes.  The cubic coordinates in terms of which the Luttinger
Hamiltonian is usually expressed are related to these coordinates by
$k_{a} = \sum_i u_{a i} k_i, ~a=x,y,z; ~i=1,2,3$
in both direct and reciprocal space.  Here the $u_{a i}$ are direction
cosines. Each element of the Luttinger Hamiltonian is a quadratic form in the
$k_i$'s. The two-dimensional hole gas is created by a GaAs (narrow gap) quantum
well flanked by the AlGaAs (wide gap) barriers. Since the barrier lies in the
range $100 - 400$ meV and the energy scale of interest is $\sim 10$ meV, we
take the barrier to be infinite. Single particle eigenspinors of the quantum
well Luttinger Hamiltonian can be expanded in the form
$\psi_{\vec{k}}(\vec{x}) = \sum_{i\alpha} c_{i\alpha} (\vec{k}) ~
  e^{i\vec{k}\cdot\vec{\rho}} \zeta_i(x_3) \chi_{\alpha}(\vec{x})$.
Here the 2D wavevector $\vec{k}=(k_1,k_2)$ is a good quantum number,
and $\zeta_i(x_3) \propto \sin(i \pi x_3/b)$ where $b$ is the well width.
Bloch functions $\chi_{\alpha}(\vec{x})$ are chosen such that when $k_1$  and $k_2$
are set to zero the Luttinger Hamiltonian takes a disgonalized form.

\begin{figure}
\epsfxsize=3.0in \epsfbox{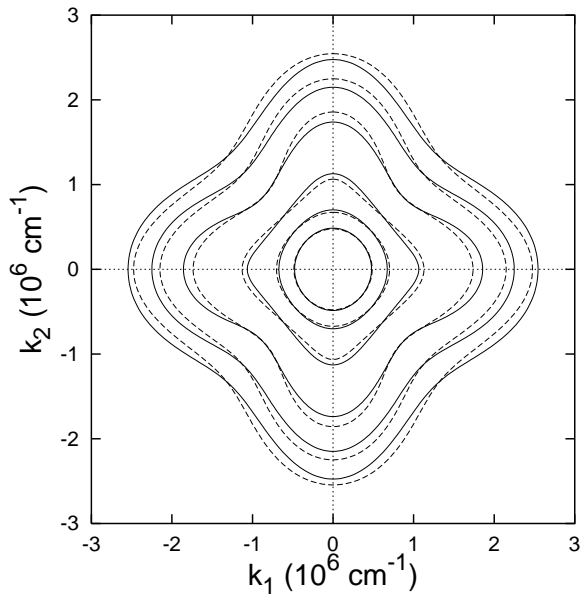} \vspace{0.5cm} \narrowtext \caption{Fermi
surface topology for the well width $250\mbox{\AA}$. The constant energy
contour lines (solid lines) are drawn for the lowest subband. $k_1$ and $k_2$
correspond to $[01\bar{1}]$ and $[\bar{2}33]$ directions, respectively. (for
illustration, rotated images of solid lines by $90^{\circ}$ are displayed as
dashed lines.) The energies for contours are $0.61, 1.1, 1.71, 2.31, 2.92$, and
$3.53$meV from inside to outside. Sample with $n_h = 1.5\times 10^{11}cm^{-2}$
corresponds to the third contour ($E_F = 1.71$meV). \label{fig.1}}
\end{figure}

For the GaAs valence bands we have used the Luttinger model parameters
$\gamma_1=6.85$, $\gamma_2=2.1$, and $\gamma_3=2.9$ and retained the first 20
subbands. A typical 2D band structure is illustrated in Fig.~\ref{fig.1} by
plotting constant energy contours for the lowest energy subband. Since our
confinement potential has inversion symmetry, each energy band is doubly
degenerate. For wider quantum wells, subband spacing is reduced and subband
mixing is strengthened.
Since cubic systems are {\em not} invariant under $90^{\circ}$ rotations about
the $[311]$ direction, the 2D bands have lower than square symmetry. The
constant energy contours are elongated in directions with larger effective
mass.  For the lowest subband, the Fermi surface anisotropy is relatively weak
near the zone center ($\Gamma$ point), but gets stronger at larger $|\vec k|$.
The effective mass is heavier (lighter) along $k_2$ than along $k_1$ direction
close to (far away from) the zone center, and is smallest along the direction
rotated from $k_1$ by $45^{\circ}$ at larger $|\vec k|$. As we shall discuss later,
the Fermi surface topology is manifested in anisotropic effective
electron-electron interactions and ultimately in orientation-dependent stripe
state energies.

When a magnetic field is applied, the subband spectrum consists of
macroscopically degenerate Landau levels whose energies may be evaluated
following familiar lines. In the Luttinger Hamiltonian, $k_1$ and $k_2$ are
replaced by raising and lowering ladder operators,
$k_1 = i(-a+a^{\dagger})/\sqrt{2}\ell$ and 
$k_2 = (a+a^{\dagger})/\sqrt{2}\ell$,
with $\ell = (\hbar c/ eB)^{1/2}$ and a Zeeman term is added to the
Luttinger Hamiltonian \cite{lutham},
$H_Z = -\kappa \mu_B \vec{J} \cdot \vec{B}$.
Here $\mu_B$ is the Bohr magneton and $\kappa=1.2$ is an additional Luttinger
model parameter. The envelope function eigenspinors can be expanded in the form
$\psi_{NX}(\vec{x})
 = \sum_{ni\alpha} c_{ni\alpha} \phi_{nX}(\vec{\rho}) \zeta_i(x_3)
  \chi_{\alpha}(\vec{x})$
where $\phi_{nX}(\vec{\rho})$ is one of the parabolic band Landau gauge
wavefunction generated by the ladder operator algebra, and the Hamiltonian
matrix is independent of the guiding center label $X$. The eigenstates of the
Hamiltonian are strong mixture of Landau level $n$ and subband $i$
indices. In diagonalizing the finite-field Luttinger Hamiltonian, three
subbands and 30 Landau levels were retained. This procedure is readily
generalized to allow for a magnetic field component perpendicular to the growth
direction, $\vec{B} = B(\hat{x}_3 + \tan\theta [\hat{x}_1 \cos\varphi +
\hat{x}_2 \sin\varphi])$. In this case, $k_1 \to k_1 + \tan\theta \sin\varphi
x_3/ \ell^2$ and $k_2 \to k_2 - \tan\theta \cos\varphi x_3/\ell^2$.

\begin{figure}
\epsfxsize=3.0in \epsfbox{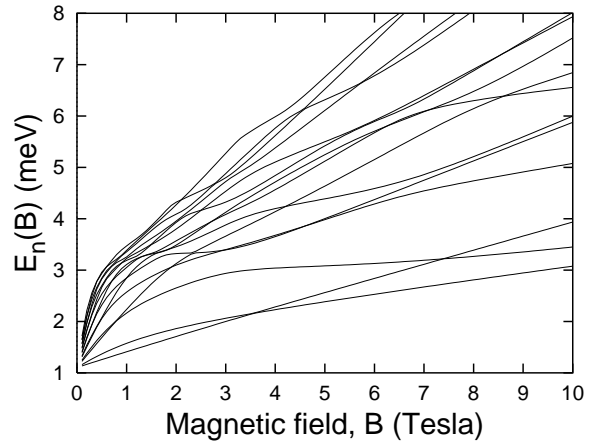} \vspace{0.5cm} \narrowtext
\caption{Landau level dispersion with magnetic field. With the inclusion of
 the Zeeman term ($\kappa=1.2$), the spin degeneracy is lifted.
\label{fig.2}}
\end{figure}

Fig.~\ref{fig.2} illustrates our results for the magnetic field dependence of
the Landau level energies for a typical quantum well width and no in-plane
field. The Landau levels show very strong nonlinear dispersion with magnetic
field.  The energy levels are unevenly spaced and the apparently crossed levels
are split due to the lack of parity under the inversion symmetry operation. 
Both Landau-level and subband mixing are stronger at larger quantum well widths.

To model the stripe state seen in Ref.~\cite{hole} at $\nu=5/2$, we consider
interacting electrons in the third level of Fig. 2 at $B\approx 2.5$ Tesla
($n_h = 1.5\times 10^{11}$ cm$^{-2}$). The distorted semiclassical cyclotron
orbit of a  Fermi energy electron with this density is illustrated in Fig.1 for
the case of $b=250\mbox{\AA}$. Semiclassical orbit distortions translate
quantum mechanically into mixing of parabolic band Landau levels.  For
interactions within a Landau level this mixing is described
exactly\cite{jungwirth} simply by replacing the Coulomb interaction by
\begin{eqnarray}
V_{\rm{eff}}(\vec{q})
 &=& {2\pi e^2 \over q\epsilon(q)}
  \sum_{\{n_i\}} M_{n_1n_4, n_2n_3} (q) ~F_{n_1n_4}(\vec{q})
   F_{n_2n_3}(-\vec{q}).
\end{eqnarray}
where $\epsilon(q)$ is the dielectric constant arising from polarization of the
lower filled Landau levels and
\begin{eqnarray}
&& M_{n_1n_4, n_2n_3} (q) \nonumber\\
 && \hspace{0.5cm} =  \sum_{\alpha\beta} \sum_{\{i_i\}}
   c_{n_1i_1\alpha}^* c_{n_4i_4\alpha}
   c_{n_2i_2\beta}^* c_{n_3i_3\beta} 
   W(i_1i_4, i_2i_3; q) \nonumber
\end{eqnarray}
with
\begin{eqnarray}
&& W(i_1i_4, i_2i_3; q)  \nonumber\\
 && \hspace{0.5cm} \equiv \int dz \int dz' ~ e^{-q|z-z'|}
   \zeta_{i_1}^* (z) \zeta_{ i_4} (z)
   \zeta_{i_2}^* (z') \zeta_{ i_3} (z'). \nonumber
\end{eqnarray}
Note that $V_{\rm{eff}}(\vec{q})$ is always repulsive.
The parabolic-band plane-wave matrix elements, $F_{nn'}(\vec{q})$, play a
crucial role since they are dependent on the 2D angular coordinate of
$\vec{q}$, $\phi_q$:
$F_{n_1n_4}(\vec{q}) F_{n_2n_3}(-\vec{q})\sim e^{i\pi(n_1+n_2-n_3-n_4)\phi_q}$.
It is precisely this effect which gives rise to orientation dependence of the
stripe state energy.  In an electron gas, $n_1=n_2=n_3=n_4=N$ and the stripe
state has no orientation dependence.

In the Hartree-Fock approximation, the energy per electron for the stripe state
of a half-filled Landau level is\cite{jungwirth}
\begin{eqnarray}
\label{ucdwen} E &=& {1\over 2\nu^*}\sum_{n=-\infty}^{\infty} \Delta_n^2 \;
 U\left(\frac{2\pi n}{a}\hat{e}\right),
\end{eqnarray}
where 
\begin{eqnarray}
\Delta_n &=& \nu^*\frac{\sin(n\nu^*\pi)}{n\nu^*\pi}, \nonumber
\end{eqnarray} 
$\hat{e}$ is the direction perpendicular to the stripes, $\nu^*$ the
filling fraction at upper Landau level, and
\begin{eqnarray} 
U(\vec{q}) 
 &=& \frac{V_{\rm{eff}}(\vec{q})}{2\pi\ell^2}
  - \int\frac{d^2p}{(2\pi)^2}\,V_{\rm{eff}}(\vec{p})\,
  e^{i(p_xq_y-p_yq_x)\ell^2}. \nonumber
\end{eqnarray} 

The evaluation of $U(\vec{q})$ is simplified by Fourier expanding the angle
dependence of $V_{\rm{eff}}(\vec{q})$ and taking advantage of inversion
symmetry in the 2D plane. The isotropic Fourier component is most dominant near
$q=0$. Higher Fourier components vanish at $q=0$, and oscillate between
positive and negative values as a function of $q$. When summed, the net result
is a strong suppression of $V_{\rm{eff}}(\vec{q})$ at $q\ell > 1.5$, which
helps to make $U(\vec{q})$ more negative  (see Figure 3) and favors stripe
states.

The results we obtain for the orientation dependence of the cohesive energy in
a $250\mbox{\AA}$ quantum well are presented in Fig.~\ref{fig.4}. The ground
state stripe orientation is tilted from the $[\bar{2}33]$ axis by
$\phi_{\rm{st}} \approx 23^{\circ}$. We studied the dependence of the ground
state $\phi$  on well width, $b$, finding that for $b=100\mbox{\AA}$ and less
$\phi_{\rm{st}} = 0^{\circ}$. On the other hand, for $b=150\mbox{\AA}$ and
$b=200\mbox{\AA}$\cite{hole}, we find $\phi_{\rm{st}} \approx 15^{\circ}$ and
$\phi_{\rm{st}} \approx 22^{\circ}$, respectively.  The difference in
cohesive energy per electron between $\phi=0$ and $\phi_{\rm{st}}$,
$|E_{coh}(0^{\circ})-E_{coh}(\phi_{\rm{st}})|$, is small: for $b=150, 200,
250\mbox{\AA}$ it is, respectively,  $0.7, 2.8, 4.5\%$ of the maximum
difference $|E_{coh}(90^{\circ})-E_{coh}(\phi_{\rm{st}})|$. For larger values of
$b=300\mbox{\AA}$, $350\mbox{\AA}$ and $400\mbox{\AA}$, we find $\phi_{\rm{st}}
\approx 45^{\circ}$. The ground state $\phi_{\rm{st}}$ changes gradually from
$0^{\circ}$ to $45^{\circ}$ with increasing quantum well width. Note that the
dependence of the ground state stripe direction on well width tracks the
well-width dependence of the Fermi surface topology.

We believe it is band anisotropy, and the orientation dependence of
stripe-state energy it leads to, that is primarily responsible for the
occurrence of hole stripe states at $\nu = 5/2$.  In 2D electron gases, stripe
states appear\cite{lilly2,rezayi} at $\nu =5/2$ only when anisotropy is imposed
by adding an in-plane magnetic field.  Comparing experiment\cite{lilly2} and
in-plane field calculations\cite{jungwirth}, we find that conduction band
stripe states appear at $\nu=5/2$ when the intrinsic orientation energy per
electron is $\sim 0.0002 e^2/\epsilon_0\ell$, comparable to the valence band 
$\nu=5/2$ orientation energy per hole at {\em perpendicular} fields 
illustrated in Fig.~\ref{fig.4}.

The valence band orientation energy is also sensitive to in-plane field as
illustrated in Fig.~\ref{fig.4} for $B\approx 2.5$ Tesla. For quantum wells of 
widths $b=100-250\mbox{\AA}$, increasing in-plane field along the $[01\bar{1}]$
axis (Fig.~\ref{fig.4}(a): $\varphi=0^{\circ}$) tends to orient the easy axis of
striped CDW closer to $[\bar{2}33]$. On the other hand, when in-plane field is
applied along $[\bar{2}33]$ axis (Fig.~\ref{fig.4}(b): $\varphi=90^{\circ}$) we find
that for a sufficiently large well widths and field tilt angles, the ground
state orientation can change from near $[\bar{2}33]$ to $[01\bar{1}]$.

In comparing with experiment, it is necessary to use the appropriate sample
width and 2D hole density and to account for the corrugations which
occur\cite{surfcorr} at GaAs/AlGaAs(311)A interfaces. For example, we have
investigated the hole density dependence of the orientation energy and found
that there is more anisotropy at higher hole densities as expected from
Fig.~\ref{fig.1}.
The consequences of corrugation for the stripe states are difficult to quantify,
partly because the stripe state period does not match the period of corrugation
potential.  Nevertheless it is likely that this morphological feature is partly
responsible\cite{hole} for the factor of $\sim 2$ mobility anisotropy at zero
field and that it will tend to align the stripes. Taking account of the
corrugations, our finding for the $200\mbox{\AA}$ quantum well case\cite{hole}, 
is consistent with experimental finding that stripes are aligned along
$[\bar{2}33]$.  It is reasonable to conclude that the corrugation contribution
to the orientation energy overcomes the small intrinsic value of
$|E_{coh}(\phi=0^{\circ}) - E_{coh}(\phi_{\rm{st}})|$.  Intrinsic orientation energies
can, however, be altered by tilted fields. For fields along the $[01\bar{1}]$
direction, $E_{coh}(\phi=90^{\circ})-E_{coh}(\phi=0^{\circ})$ is increased and we would
expect little experimental consequence.  For fields along the $[\bar{2}33]$
direction, on the other hand, we predict a dramatic reversal of stripe
orientations at tilt angle $\theta \sim 60^{\circ}$. At this angle, the
intrinsic band-structure effects addressed here favor $[01\bar{1}]$ oriented
stripes by $\sim 0.0008 e^2/\epsilon_0\ell$ per electron. This number should be 
compared with the $\sim 0.0001 e^2/\epsilon_0\ell$ orientation energy 
produced\cite{jungwirth} by a
$25^{\circ}$ field tilt in $n=2$ conduction band Landau levels. In the $[100]$
growth conduction band case, stripes orient along $[110]$ directions for
perpendicular fields, likely because of MBE growth stabilities analogous to the
corrugations discussed above. The anisotropy energy produced by a $~\sim
25^{\circ}$ field tilt, in the $n=2$ conduction band case is sufficient to
overcome this extrinsic anisotropy and reorient the stripes.  Based on these
comparisons, we conclude that tilted field effects can also overcome extrinsic
anisotropy sources in the valence band case.

In summary, valence band stripe states have a finite orientation energy even
without an in-plane field, favoring their occurrence at $\nu=5/2$.  For quantum
well widths $b = 100 - 250\mbox{\AA}$, intrinsic band effects yield a ground
state orientation close to $[\bar{2}33]$ and corrugation effects are likely
sufficient to produce the $[\bar{2}33]$ orientation seen in experiment. We
predict that for typical quantum well widths, an in-plane field applied along
the $[\bar{2}33]$ direction will induce a stripe reorientation transition.

 We are indebted to J.P. Eisenstein, T. Jungwirth, M. Shayegan, and R. Winkler for
valuable discussions.  This work was supported in part by the National Science
Foundation under grant DMR-9714055, and in part by 
grant No.1999-2-112-001-5 from the interdisciplinary Research program of the KOSEF.
We thank the Center for Theoretical Physics
at Seoul National University for providing computing time.

\begin{figure}
\epsfxsize=5.0in \noindent
\begin{minipage}[t]{.49\linewidth}
 \centering\epsfig{file=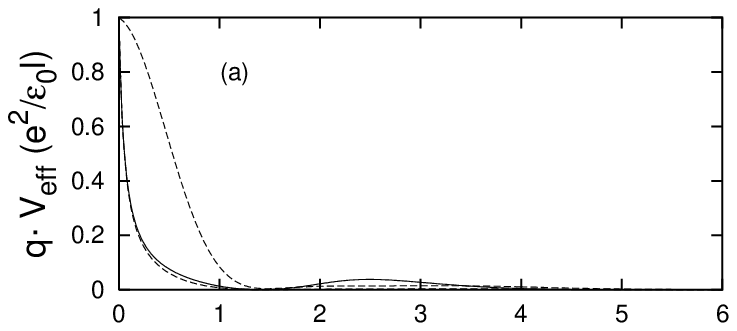,width=7.cm}
\end{minipage}

\noindent
\begin{minipage}[t]{.49\linewidth}
 \centering\epsfig{file=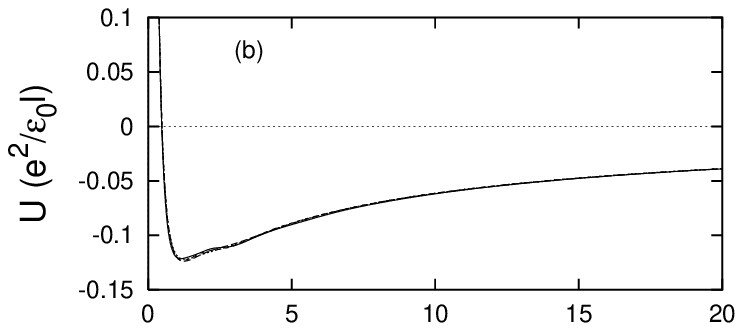,width=7.25cm}
\end{minipage}

\noindent
\begin{minipage}[t]{.49\linewidth}
 \centering\epsfig{file=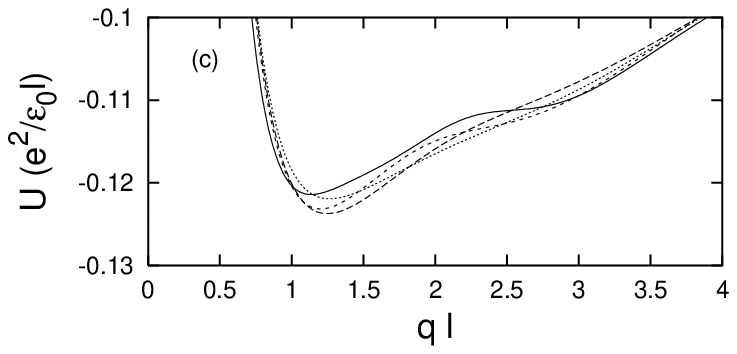,width=7.5cm}
\end{minipage}
\vspace{0.5cm} \narrowtext \caption{Comparison of effective Coulomb potentials.
(a) The Hartree potentials along $[01\bar{1}]$ are displayed for the quantum
well of width $250\mbox{\AA}$. Solid line corresponds to no-Landau-level-mixing
case with screening. The upper (lower) dashed line represents unscreened
(screened) Hartree potential for 2D holes at $\nu=5/2$. (b) Hartree-Fock (HF)
potentials. (c) Magnified view of HF potentials near the minimum where the
angle dependence is significant. In (b) and (c), solid line: without
Landau-level-mixing; long dashed line: along $[01\bar{1}]$; short dashed line:
along $45^{\circ}$ from $x_1$ axis; dotted line: along $[\bar{2}33]$.
\label{fig.3}}
\end{figure}

\begin{figure}
\epsfxsize=3.0in \noindent
\begin{minipage}[b]{.49\linewidth}
 \centering\epsfig{file=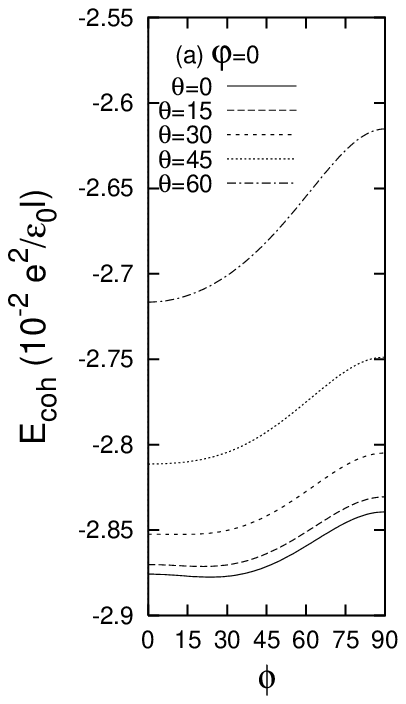,width=\linewidth}
\end{minipage}
\begin{minipage}[b]{.49\linewidth}
 \centering\epsfig{file=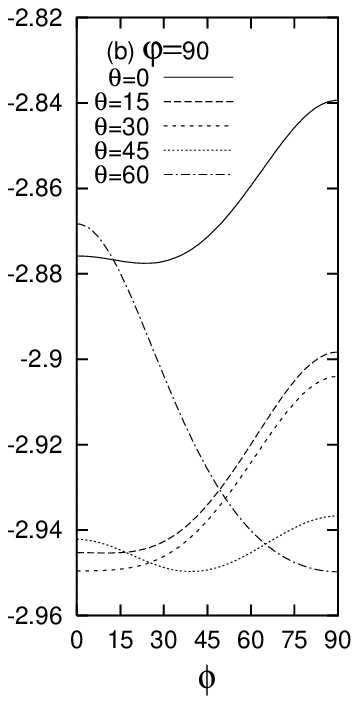,width=\linewidth}
\end{minipage}
\vspace{0.5cm} \narrowtext \caption{
 The cohesive energy at $\nu=5/2$ is drawn as a function of $\phi$
 (the angle between $[\bar{2}33]$ and the easy axis of the striped CDW)
 with varying in-plane magnetic field
 for the confinement potential width $250\mbox{\AA}$.
 Solid lines correspond to the case without in-plane fields.
 $\theta$ defines the direction of tilted magnetic fields from the $[311]$
 axis. The in-plane fields are pointing along
 $[01\bar{1}]$ and $[\bar{2}33]$ in (a) and (b), respectively.
\label{fig.4}}
\end{figure}

\begin{thebibliography}{100}

\bibitem{els1} M. P. Lilly, K.B. Cooper, J.P. Eisenstein, L.N. Pfeiffer, and K.W.
West, Phys. Rev. Lett. {\bf 82}, 394 (1999).

\bibitem{du} R.R. Du, D.C. Tsui, H.L. Stormer, L.N. Pfeiffer, K.W. Baldwin, and
K.W. West, Solid State Commun. {\bf 109}, 389 (1999).

\bibitem{els2} W. Pan, R.R. Du, H.L. Stormer, D.C. Tsui, L.N. Pfeiffer, K.W.
Baldwin, and K.W. West, Phys. Rev. Lett. {\bf 83}, 820 (1999).

\bibitem{lilly2} M.P. Lilly, K.B. Cooper, J.P. Eisenstein, L.N. Pfeiffer, and
K.W. West, Phys. Rev. Lett. {\bf 83}, 824 (1999).

\bibitem{cdw} A.A. Koulakov, M.M. Fogler, and B.I. Shklovskii, Phys. Rev. Lett.
{\bf 76}, 499 (1996); Phys. Rev. B {\bf 54}, 1853 (1996); R. Moessner and J. T.
Chalker, Phys.\ Rev.\ B {\bf 54}, 5006 (1996).

\bibitem{leshouches} See for example A.H. MacDonald in {\it Proceedings of the
Les Houches Summer School on Mesoscopic Physics} (Elsevier,Amsterdam,1995),
edited by  E. Akkermans, G. Montambeaux, and J.-L. Pichard.

\bibitem{rezayi} E.H. Rezayi, F.D.M. Haldane, and Kun Yang, Phys. Rev.
Lett. {\bf 83}, 1219 (1999).

\bibitem{jpe00} J.P. Eisenstein, M.P. Lilly, K.B. Cooper, L.N. Pfeiffer, and K.W.
West, preprint [cond-mat/0003405], to appear in Physica E.

\bibitem{fertig} H.A. Fertig, Phys. Rev. Lett. {\bf 82}, 3693 (1999).

\bibitem{kivelson} E. Fradkin and S.A. Kivelson, Phys. Rev. B {\bf 59}, 8065 (1999).

\bibitem{ahm} A.H. MacDonald, and Matthew P.A. Fisher, Phys. Rev. B {\bf 61}, 5724
(2000).

\bibitem{jungwirth} T. Jungwirth, A.H. MacDonald, L. Smrcka, and S.M. Girvin,
Phys. Rev. B {\bf 60}, 15574 (1999).

\bibitem{phillips} T. Stanescu, I. Martin, and P. Phillips,
Phys. Rev. Lett. {\bf 84}, 1288 (2000).

\bibitem{pan} W. Pan, T. Jungwirth, H.L. Stormer, D.C. Tsui, A.H. MacDonald,
S.M. Girvin, L. Smrcka, L.N. Pfeiffer, K.W. Baldwin, and K.W. West, preprint
[cond-mat/0003483].

\bibitem{hole} M. Shayegan, H.C. Manoharan, S.J. Papadakis, and
E.P. De Poortere, Physica E {\bf 6}, 40 (2000).  The sample studied in this
experimental sample had a quantum well width $b = 200\mbox{\AA}$.

\bibitem{lutham} J.M. Luttinger, Phys. Rev. {\bf 102}, 1030 (1956).



\bibitem{surfcorr} R. N\"otzel, N.N. Ledentsov, L. D\"aweritz, K. Ploog, and M.
Hohenstein, Phys. Rev. B {\bf 45}, 3507 (1992).

\end{thebibliography}
\end{document}